\documentclass[pra,10pt,aps,showpacs]{revtex4}

\usepackage{amssymb}
\usepackage{amsmath}
\usepackage{graphicx}
\usepackage{dcolumn}
\usepackage{bm}
\usepackage{epstopdf}

\begin{document}
\title{Single-point position and transition defects in continuous time
quantum walks}
\author{Zhi-Jian Li$^{1,\ast}$ \& J. B. Wang$^{2,\dag}$}
\affiliation{{$^{1}$Institute of Theoretical Physics, Shanxi University, Taiyuan, 030006,
China}\\
{$^{2}$School of Physics, The University of Western Australia, WA 6009,
Australia}\\
{$^\ast$ Email: zjli@sxu.edu.cn and $^\dag$Email: jingbo.wang@uwa.edu.au}}
\begin{abstract}
We present a detailed analysis of continuous time quantum walks (CTQW) with
both position and transition defects defined at a single point in the line.
Analytical solutions of both traveling waves and bound states are obtained,
which provide valuable insight into the dynamics of CTQW. The number of
bound states is found to be critically dependent on the defect parameters,
and the localized probability peaks can be readily obtained by projecting
the state vector of CTQW on to these bound states. The interference between
two bound states are also observed in the case of a transition defect. The
spreading of CTQW probability over the line can be finely tuned by varying
the position and transition defect parameters, offering the possibility of
precision quantum control of the system.
\end{abstract}
\maketitle
\section{Introduction}
Compared to the classical random walk, which is a memoryless Markov process,
a quantum walk is unitary and time-reversible~\cite{discrete,continuous}. It
exhibits markedly different behavior due to superposition, interference, and
quantum correlations. For instance, a quantum walk can propagate
quadratically faster than its classical counterpart and result in a
probability distribution vastly different from the classically expected
Gaussian distribution~\cite{kempe}. Quantum walks have become useful tools
for modeling and analyzing the behavior of quantum systems, for simulating
biological processes such as energy transfer in photosynthesis~\cite%
{photosynthesis}, for studying quantum phenomena such as perfect state
transfer~\cite{statetransfer}, Anderson localization~\cite{localization} and
topological phases~\cite{topology}, as well as for developing novel quantum
algorithms in quantum information processing~\cite{algorithms,childs}.
Experimentally, quantum walks have been implemented in a variety of systems,
such as nuclear magnetic resonance~\cite{nmr}, trapped ions and trapped cold
neutral atoms ~\cite{ion,atom}, single photons in bulk ~\cite{photon}, fiber
optics~\cite{fiber}, and coupled waveguide arrays~\cite{waveguide}.

With the physical implementation of quantum walks comes the issue of
disorder and decoherence. The effects of decoherence and disorder on the
quantum walks have been extensively studied, for example, their transition
to classical random walks under the influence of decoherence~\cite%
{dec1,dec2,dec3}. Static and dynamic disorder also alters quantum walks from
ballistic spread to localization through a disruption of the interference
pattern~\cite{dis1,dis2,dis3,dis4,dis5,dis6}. Recently, W\'{o}jcik et al.~%
\cite{wojcik}, Li et al.~\cite{we1} and Zhang et al.~\cite{xue} investigated
the localization property of one-dimensional discrete time quantum walks
(DTQW) with a single-point phase defect. Motes et al.~\cite{motes} use a
bit-flip coin at a boundary to introduce the position defects and find the
walker escapes dramatically faster through the boundary. For continuous time
quantum walk with defects, although a precursor work by Koster and Slater~%
\cite{koster} has explored quantitatively the limiting case of a single
diagonal defect in a one-dimensional molecular crystal using a
nearest-neighbor tight-binding model, an analytic derivation is absent to
provide insight for the prevalence results relying on numerical methods.
Li et al. \cite{we1} and Izaac et al.\cite{we2} have compared similar behaviors between CTQWs
and DTQWs with single- and double-point defects. In this paper, we extend
these works to include not only position defects but also transition defects
in continuous time quantum walks, presenting analytical solutions of both
traveling waves or bound states of CTQWs in position space. Here, the bound
state means that the quantum walk is localized in one region of the position
space with zero probability in the limit of asymptotic infinity. We use its
analytical expression to discuss the associated eigenstate localization.

\section{Results}

\subsection{The single-point defect model of CTQW}

The continuous time quantum walk was first posited by Farhi and Gutmann~\cite%
{continuous}, as a quantization of the corresponding classical continuous
time random walk. In CTQWs, classical probabilities are replaced by quantum
probability amplitudes, with the system evolving as per the Schr\"{o}dinger
equation in discrete space, rather than the Markovian master equation~\cite%
{wang}. To illustrate, we consider a classical continuous time random walk
on the discrete graph $G(V$,$E)$ described by two sets $V$ and $E$. The set $%
V$ is composed of the unordered nodes $j$ and the set $E$ includes the edges 
$e_{jk}=(j,k)$ connecting the node $j$ to the node $k$. The transition rate
matrix $H$ is defined as 
\begin{equation}
\begin{array}{c}
H_{jk}=\left\{ 
\begin{array}{ll}
\gamma _{jk} & \text{for $j\neq k$ and }e_{jk}\in E \\ 
0 & \text{for $j\neq k$ and }e_{jk}\notin E \\ 
-\varepsilon _{j} & \text{for $j=k$}%
\end{array}%
\right.%
\end{array}
\label{eq:h0}
\end{equation}%
where $\gamma _{jk}$ is the probability per unit time for making a
transition from node $k$ to node $j$. For the probability to be
conservative, the constraint 
\begin{equation}
\varepsilon _{j}=\sum_{k,k\neq j}^{N}\gamma _{jk},
\label{eqn.trans-mat-conservative}
\end{equation}%
is required, where $N$ is the total number of nodes in the graph. If the
transition rates between any two connected nodes are the same, i.e. $\gamma
_{jk}=\gamma $, the diagonal element $\varepsilon _{j}=d_{j}\gamma $ with $%
d_{j}$ denoting the degree of the node $j$ or the number of sites connected
to node $j$. The state of the random walker is fully described by the
probability distribution vector $\mathbf{P}(t)$, with its time evolution
governed by the master equation 
\begin{equation}
\frac{d\mathbf{P}(t)}{dt}=H\mathbf{P}(t),
\end{equation}%
which has the formal solution $\mathbf{P}(t)=e^{Ht}\mathbf{P}(0)$.

Extending the above description to the quantum realm involves replacing the
real valued probability distribution vector $\mathbf{P}(t)$ with a complex
valued wave function $|\psi (t)\rangle$ and adding the complex notation $-i $
to the evolution exponent, namely 
\begin{equation}
|\psi (t)\rangle =e^{-iHt}|\psi (0)\rangle .  \label{eq:se}
\end{equation}%
The quantum transition matrix $H$, often referred to as the system
Hamiltonian, is required to be Hermitian instead of being constrained by
Eq.~(\ref{eqn.trans-mat-conservative}). Consequently, the above time
evolution is unitary, guaranteeing that the norm of $|\psi (t)\rangle $ is
conserved under a CTQW. Let $\mathbf{j}$ be the position operator with
eigenvector $|j\rangle$. The system state vector can be expanded in the
position Hilbert space with basis $\left\{ |j\rangle \right\}$, $|\psi
(t)\rangle =\sum_{j}a_{j}(t)|j\rangle$ where $a_{j}(t)=\langle j|\psi
(t)\rangle $ represents the probability amplitude of the walker being found
at node $j$ at time $t$. The resulting probability distribution is given by $%
P_{j}=|a_{j}(t)|^{2}=|\langle j|\psi (t)\rangle |^{2}$.

For a CTQW on a uniform infinite line, its Hamiltonian can be expressed as%
\begin{equation}
H_{0}=\varepsilon \sum\limits_{j}\left\vert j\right\rangle \left\langle
j\right\vert -\gamma \sum\limits_{j}\left( \left\vert j+1\right\rangle
\left\langle j\right\vert +\left\vert j-1\right\rangle \left\langle
j\right\vert \right) .
\end{equation}%
Here, each node is connected to its neighboring nodes by a constant
transition rate $\gamma $, and each node has a constant potential energy $%
\varepsilon $. Now we introduce two types of single-point defects in this
model, one being a position defect that has a different potential energy $%
\alpha $ at node $j_{d}$ and the other as a transition defect, where a
distinctive transition rate $\beta $ is assigned. Without loss of
generality, we assume that the parameters $\varepsilon$, $\gamma$ , $\alpha$
and $\beta$ are reals. To account for these defects, the system Hamiltonian
is modified as 
\begin{equation}
H=H_{0}+H_{1}+H_{2},
\end{equation}
with 
\begin{eqnarray}
H_{1} &=&\alpha |j_{d}\rangle \langle j_{d}|, \\
H_{2} &=&-\beta \left( |j_{d}\rangle \langle j_{d}+1|+|j_{d}+1\rangle
\langle j_{d}|+|j_{d}\rangle \langle |j_{d}-1|+|j_{d}-1\rangle \langle
j_{d}|\right) .
\end{eqnarray}%
The position energy at the defect node $j_{d}$ is $\varepsilon +\alpha $ and
the transition rate between it and its neighboring nodes is $\gamma +\beta $.

\subsection{Eigen problem of the model Hamiltonian}

%\subsubsection*{The parity property}
The Hamiltonian of the CTQW on an uniform infinite line is invariant under
spatial translation. Consider the discrete translational operator $\mathbf{T}%
_{n}$, which acts on the node states such that $\mathbf{T}_{n}|j\rangle
=|j+n\rangle $. This operator is unitary, and as such can be written in the
form $\mathbf{T}_{n}=e^{i\mathbf{k}n} $, where $\mathbf{k}$ is an Hermitian
operator and the generator of the translation. In the case where the
Hamiltonian is invariant under spatial translation, the Hermiticity of $%
\mathbf{k}$ indicates that its eigenstates $|k\rangle
=\sum_{j}e^{ikj}|j\rangle $ form a complete orthonormal basis, satisfying
the eigenvalue equation $H_{0}|k\rangle =\left( \varepsilon -2\gamma \cos {k}%
\right) |k\rangle $, where $0\leq k\leq \pi $. The addition of a defect
breaks the translational symmetry of the system, which results in an
emergence of localized eigenstates of the corresponding quantum walk. The
eigenstates of CTQW on a infinite line with a single-point defect can be
obtained by solving a set of recurrence equations as the following.

Expanding the eigenstate $\left\vert \psi \right\rangle $ of $H$ in the
position space as $\left\vert \psi \right\rangle
=\sum\limits_{j}C_{j}\left\vert j\right\rangle$ and substituting it into the
eigen equation $\left\langle j\right\vert H\left\vert \psi \right\rangle
=\lambda \left\langle j|\psi \right\rangle$ with eigenvalue $\lambda $, we
get a set of recurrence equations about $C_{j}$ 
\begin{eqnarray}
&&\gamma\text{ }C_{j+1}-\left( \varepsilon -\lambda \right) C_{j}+\gamma
C_{j-1} =0\text{ \ \ \ \ \ \ \ for }j\neq j_{d},\text{ }j_{d}\pm 1~,
\label{Cj} \\
&&\left( \gamma +\beta \right) C_{j_{d}+1}+\left( \gamma +\beta \right)
C_{j_{d}-1}-\left( \varepsilon +\alpha -\lambda \right) C_{j_{d}} =0~, \text{
\ \ \ \ \ \ }  \label{cz1} \\
&&\left( \gamma +\beta \right) C_{j_{d}}+\gamma C_{j_{d}+2}-\left(
\varepsilon -\lambda \right) C_{j_{d}+1} =0~,\text{ }  \label{c0} \\
&&\left( \gamma +\beta \right) C_{j_{d}}+\gamma C_{j_{d}-2}-\left(
\varepsilon -\lambda \right) C_{j_{d}-1} =0~.  \label{cf1}
\end{eqnarray}
The general solution of Eq.~(\ref{Cj}) is 
\begin{equation}
C_{j}=Ay^{(j-j_{d})}+By^{-(j-j_{d})}\text{ \ \ \ \ \ for }j\neq j_{d},\text{ 
}j_{d}\pm 1~,  \label{gs}
\end{equation}%
where $A$ and $B$ are arbitrary constants, and $y$ satisfies the following
equation 
\begin{equation}
\left( y-\frac{\varepsilon -\lambda }{\gamma }-\frac{1}{y}\right) =0
\label{yeq}
\end{equation}%
Solving the above equation yields $y_{\pm }=\frac{\left( \varepsilon
-\lambda \right) \pm \sqrt{-4\gamma ^{2}+\left( \varepsilon -\lambda \right)
^{2}}}{2\gamma }~.$ It can be easily shown that $y_{+}=y_{-}^{-1}$, and
therefore we only need to substitute $y=y_{+}$ into Eq.~(\ref{gs}) as our
general solution.

Due to the reflection symmetry of the underlying potential with defects at a
single node $j=j_{d}$, the system eigenvectors in position space must possess either an odd
or even parity at the defect node. In the case of odd parity, i.e. $%
C_{j}=-C_{-j+2j_{d}}$, we let $C_{j}=sign(j-j_{d})\left( Ay^{\left\vert
j-j_{d}\right\vert }+By^{-\left\vert j-j_{d}\right\vert }\right) $,
Substituting this into Eqs.(\ref{Cj}-\ref{cf1}) and using Eq.(\ref{yeq}), we
obtain the coefficients as 
\begin{equation}
B=-A,\text{ \ }C_{j_{d}}=0,\text{ \ and ~}C_{j_{d}+1}=-C_{j_{d}-1}=A\frac{%
\sqrt{-4\gamma ^{2}+(\varepsilon -\lambda )^{2}}}{\gamma }~.  \label{codd}
\end{equation}%
In the case of even parity, i.e. $C_{j}=C_{-j+2j_{d}}$, we let $%
C_{j}=Ay^{\left\vert j-j_{d}\right\vert }+By^{-\left\vert j-j_{d}\right\vert
}$ and the coefficients are 
\begin{equation}
B=f(\lambda )A,\text{ \ }C_{j_{d}}=\frac{\gamma }{\gamma +\beta }%
(1+f(\lambda ))A,\text{ \ }C_{j_{d}+1}=C_{j_{d}-1}=\frac{\gamma (\alpha
+\varepsilon -\lambda )}{2(\gamma +\beta )^{2}}(1+f(\lambda ))A~,
\label{ceven}
\end{equation}%
where%
\begin{equation}
f(\lambda )=\frac{-\left( \alpha +\varepsilon -\lambda \right) \gamma
^{2}+\left( \gamma +\beta \right) ^{2}\left( \varepsilon -\lambda +\sqrt{%
-4\gamma ^{2}+\left( \varepsilon -\lambda \right) ^{2}}\right) }{\left(
\alpha +\varepsilon -\lambda \right) \gamma ^{2}-\left( \gamma +\beta
\right) ^{2}\left( \varepsilon -\lambda -\sqrt{-4\gamma ^{2}+\left(
\varepsilon -\lambda \right) ^{2}}\right) }~.  \label{fffff}
\end{equation}
The arbitrary constant A in Eqs. (\ref{codd}) and (\ref{ceven}) will be
determined by the normalized condition of the state vector.

%\subsubsection*{Traveling-wave solution}
The eigen vectors are traveling waves or bound states in position space are
modulated by the module value of y, which depends on the eigenvalues of the
system. When $\lambda \in \left[ \varepsilon -2\left\vert \gamma \right\vert
,\varepsilon +2\left\vert \gamma \right\vert \right]$, $\left\vert
y\right\vert =1$ and we can set $y=e^{ik}.$ The solution given by Eq.~(\ref%
{gs}) is thus a traveling wave, and the corresponding eigenvalue $\lambda $
can be obtained from Eq.~(\ref{yeq}) 
\begin{equation}
\lambda =\lambda _{k}=\varepsilon -2\gamma \cos(k)~,  \label{tev}
\end{equation}%
where $k$ is analogous to the wave number of free particle in period
lattice. Substituting Eq.~(\ref{tev}) into Eq.~(\ref{codd}) and Eq.(\ref%
{ceven}) , we get$\ $the normalized odd-parity traveling eigenvector 
\begin{equation}
\left\vert \psi _{k}^{o}\right\rangle =\frac{i}{\sqrt{\pi }}%
\sum\limits_{j}\sin \left[ k\left( j-j_{d}\right) \right] \left\vert
j\right\rangle ~.  \label{evodd}
\end{equation}%
and even-parity traveling eigenvector 
\begin{eqnarray}
\left\vert \psi _{k}^{e}\right\rangle &=&\frac{1+f(\lambda _{k})}{\sqrt{4\pi
-\left[ 2+f(\lambda _{k})+f^{\ast }(\lambda _{k})\right] \left[ 1-\left( 
\frac{\gamma }{\gamma +\beta }\right) ^{2}\right] }}  \notag \\
&&\left[ \frac{-\beta }{\gamma +\beta }\left\vert j_{d}\right\rangle
+\sum\limits_{j}\left( \cos (k\left\vert j-j_{d}\right\vert )+i\frac{%
1-f(\lambda _{k})}{1+f(\lambda _{k})}\sin (k\left\vert j-j_{d}\right\vert
)\right) \left\vert j\right\rangle \right] ~,  \label{eveven}
\end{eqnarray}%
respectively, where 
\begin{equation}
f(\lambda _{k})=\frac{2i\left( \gamma +\beta \right) ^{2}\sin (k)-\left[
\gamma \alpha -2\beta (2\gamma +\beta )\cos (k)\right] }{2i\left( \gamma
+\beta \right) ^{2}\sin (k)+\left[ \gamma \alpha -2\beta (2\gamma +\beta
)\cos (k)\right] }~.  \label{fff}
\end{equation}%
We note that odd-parity traveling eigenvector is independent on the defect
parameters $\alpha$ and $\beta$, just like on the uniform infinite lattice
line traveling with constant amplitude. It is very different for the
even-parity traveling eigenvector, in which the wave traveling towards right
and the wave traveling towards left have different amplitudes and they are
inversion symmetry about the defect position. The amplitudes are adjusted
not only by the defect parameters but also by the wave number $k$. If only $%
\beta =0$, the the even-parity traveling eigenvector reduces to $\left\vert
\psi _{k}^{e}\right\rangle =\frac{1}{2\sqrt{\pi }}\frac{4\gamma \sin (k)}{%
-i\alpha +2\gamma \sin (k)}\sum_{j}\left[ \cos (k\left\vert
j-j_{d}\right\vert )+\frac{\alpha \csc (k)}{2\gamma }\sin (k\left\vert
j-j_{d}\right\vert )\right] \left\vert j\right\rangle $ as given by Izaac et
al.~\cite{we2}. If both $\beta =0$ and $\alpha =0$, it comes back to the
free case $\left\vert \psi _{k}^{e}\right\rangle =\frac{1}{\sqrt{\pi }}%
\sum_{j}\cos (k\left\vert j-j_{d}\right\vert )\left\vert j\right\rangle .$

%\subsubsection*{Bound state solution}
When $\lambda <\varepsilon -2\left\vert \gamma \right\vert,$ we have $%
\left\vert y\right\vert >1$, and when $\lambda >\varepsilon +2\left\vert
\gamma \right\vert ,$ $\left\vert y\right\vert <1$. For Eq.(\ref{gs}) being
convergent at the infinity, either $A$ or $B$ must be zero. In the case of
odd-parity, there is no physical solution for $C_{j}$ due to the requirement 
$B=-A$. However, for the case of even parity, if $f^{sign(1-\left\vert
y\right\vert )}(\lambda )=0$, Eq.~(\ref{gs}) can be reduced to $%
C_{j}=Ay^{sign(1-\left\vert y\right\vert )\left\vert j-j_{d}\right\vert }$ \ 
$(j\neq j_{d}$, $j_{d}\pm 1)$, the bound eigenvector exists, and the
corresponding bound eigenvalues $\lambda _{b}$ can be obtained from solving
the equation $f^{sign(1-\left\vert y\right\vert )}(\lambda )=0$ as 
\begin{equation}
\lambda _{b}=\lambda _{\pm }=\varepsilon +\frac{\beta (2\gamma +\beta
)\alpha }{(\gamma +2\beta )^{2}-2\beta ^{2}}\pm \frac{(\gamma +\beta )^{2}}{%
(\gamma +2\beta )^{2}-2\beta ^{2}}\sqrt{4(\gamma +2\beta )^{2}-8\beta
^{2}+\alpha ^{2}}~.  \label{eqbound}
\end{equation}%
In this case the system has zero, one, or two bound eigenstates, dependent
on the value range of the parameters $\varepsilon $, $\gamma $, $\alpha $
and $\beta $ to satisfy with $\left\vert y\right\vert >1$ or $\left\vert
y\right\vert <1$. Other coefficients in Eq.~(\ref{ceven}) are found to be $%
C_{j_{d}}=\frac{\gamma }{\gamma +\beta }A$ and $C_{j_{d}+1}=C_{j_{d}-1}=%
\frac{\gamma (\alpha +\varepsilon -\lambda _{b})}{2(\gamma +\beta )^{2}}A.$
Finally, the normalized bound eigenvector with even parity can be written as 
\begin{equation}
\left\vert \psi ^{b}\right\rangle =A_{b}\left[ \sum\limits_{j\neq j_{d},%
\text{ }j_{d}\pm 1}y^{sign(1-\left\vert y\right\vert )\left\vert
j-j_{d}\right\vert }\left\vert j\right\rangle +\frac{\gamma }{\gamma +\beta }%
\left\vert j_{d}\right\rangle +\frac{\gamma (\alpha +\varepsilon -\lambda
_{b})}{2(\gamma +\beta )^{2}}\left( \left\vert j_{d}+1\right\rangle
+\left\vert j_{d}-1\right\rangle \right) \right]  \label{evbound}
\end{equation}%
with 
\begin{equation*}
A_{b}=\left[{2\frac{y^{sign(1-\left\vert y\right\vert )4}}{%
1-y^{sign(1-\left\vert y\right\vert )2}}+\left( \frac{\gamma }{\gamma +\beta 
}\right) ^{2}+2\left( \frac{\gamma (\alpha +\varepsilon -\lambda _{b})}{%
2(\gamma +\beta )^{2}}\right) ^{2}}\right]^{-1/2}.
\end{equation*}
Its distribution on the position space is centered at the defect node, and
exponentially decays with increasing of the distance from defect node. The
height of the center peak and the decaying rate are determined by the
strength of the defect.

Using the orthogonality relations of the sine and cosine functions, it can
be easily shown that %\begin{equation}
$\left\langle \psi _{k}^{o}|\psi _{k}^{e}\right\rangle =0,\left\langle \psi
_{k}^{o}|\psi ^{b}\right\rangle =0,\left\langle \psi _{k}^{e}|\psi
^{b}\right\rangle =0$ %\end{equation}%
for all values of $0\leq k\leq \pi ,$ and %\begin{equation}
$I=\sum_{b}\left\vert \psi ^{b}\right\rangle \left\langle \psi
^{b}\right\vert+\int_{0}^{\pi }dk\left( \left\vert \psi
_{k}^{o}\right\rangle \left\langle \psi _{k}^{o}\right\vert +\left\vert \psi
_{k}^{e}\right\rangle \left\langle \psi _{k}^{e}\right\vert \right).$ 
%\end{equation}%
That is to say, the eigenvectors obtained above remain orthonormal with
respect to each other and they form a complete set of basis. Consequently,
the time-evolution of an arbitrary initial state $\left\vert \psi
(0)\right\rangle $ can be constructed in an integral form as 
\begin{equation}
\left\vert \psi (t)\right\rangle =e^{-iHt}\left\vert \psi (0)\right\rangle = 
\left[ \int_{0}^{\pi }dke^{-i\lambda _{k}t}\left( \left\vert \psi
_{k}^{o}\right\rangle \left\langle \psi _{k}^{o}\right\vert +\left\vert \psi
_{k}^{e}\right\rangle \left\langle \psi _{k}^{e}\right\vert \right)
+\sum_{b}e^{-i\lambda _{b}t}\left\vert \psi ^{b}\right\rangle \left\langle
\psi ^{b}\right\vert \right] \left\vert \psi (0)\right\rangle .
\label{timeevolution}
\end{equation}
We have verified numerically in the following calculation that the integral
result given by the above equation is completely consistent with that
obtained by taking the matrix exponential of the Hamiltonian directly from
Eq.~(\ref{eq:se}).

\subsection{The effect of a position defect}

Choosing the parameter values $\varepsilon =2, \gamma =1$ and $\beta =0$,
we firstly examine the effects of a position defect on the quantum walk. In
this case, there is always one bound state as long as $\alpha\neq 0$. The
bound eigen energy $\lambda _{b}$ as a function of $\alpha$ is shown in
Fig.1, in which $\lambda _{b}=\lambda _{+}>\varepsilon +2\gamma$ if $\alpha
>0$ or $\lambda _{b}=\lambda _{-}<\varepsilon -2\gamma$ if $\alpha <0$. %
\begin{figure}[th]
\begin{center}
\includegraphics [width=10cm] {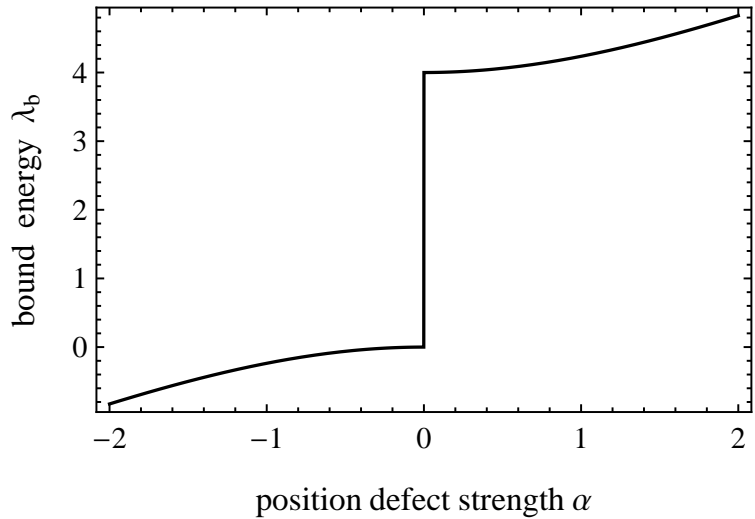}
\end{center}
\caption{The variation of bound energy with the strength of position defect.}
\end{figure}

The left panel of Fig.2 shows the CTQW probability distribution at $t=30$,
given that the quantum walk initially starts at the origin $j_{0}=0$, the
strength of defect $\alpha =3$, and the defect position $j_{d}=0,1,2,5$,
respectively. % 
\begin{figure}[th]
\begin{center}
\includegraphics [width=16cm] {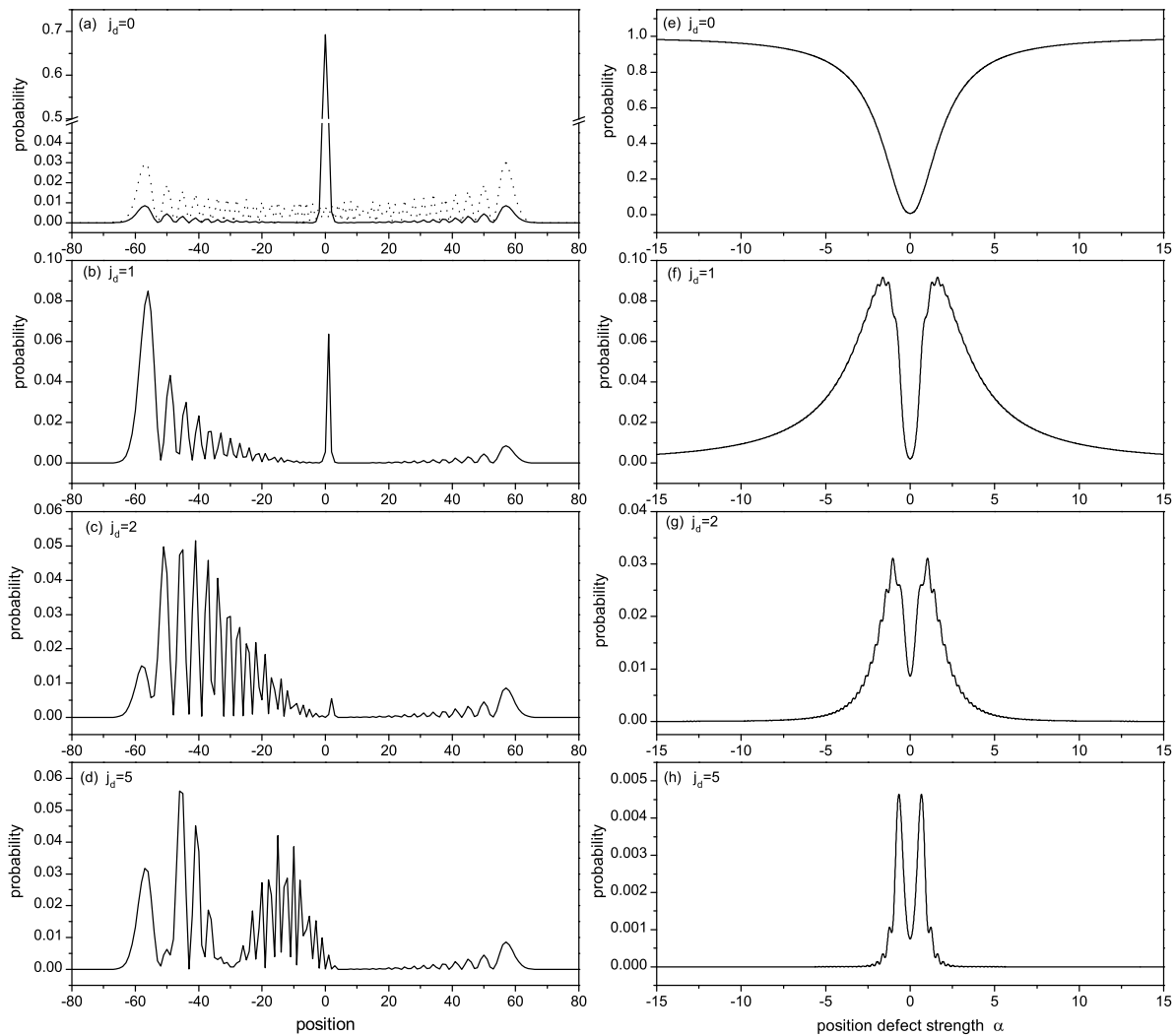}
\end{center}
\caption{ Left panel: the probability distribution of CTQW with a
single-point position defect when $t=30$, $\alpha=3$, $j_{0}=0$ and $%
j_{d}=0,1,2,5$; Right panel: the probability at the defect position as a
function of position defect strength.}
\end{figure}
If a defect is located at the initial position $j_{d}=j_{0}$, a large sharp
peak appears at this position (see Fig.2(a)) and its height remains largely
unchanged with time. For comparison, the dashed line depicts the probability
distribution of the free quantum walk without the defect. When the defect
position is the nearest to the initial position of CTQW, i.e. $\left\vert
j_{d}-j_{0}\right\vert =1$, the probability distribution also has a small
peak localized at the defect position (see Fig.2(b)). However, when the
defect position deviates away the initial position more a little, i.e. $\left\vert
j_{d}-j_{0}\right\vert >1$, the CTQW probability at the defect position
decrease rapidly to a minimum (see Fig.2(c)(d)). This phenomenon is related
to the bound state induced by the presence of a single defect. It can be
readily illustrated through decomposed form of the CTQW probability at the
defect position
\begin{eqnarray}
P_{j_{d}}=\left\vert \int_{0}^{\pi }dke^{-i\lambda _{k}t}\left\langle j_{d}|\psi
_{k}^{o}\right\rangle \left\langle \psi _{k}^{o}|j_{0}\right\rangle
+\int_{0}^{\pi }dke^{-i\lambda _{k}t}\left\langle j_{d}|\psi
_{k}^{e}\right\rangle \left\langle \psi _{k}^{e}|j_{0}\right\rangle
+e^{-i\lambda _{b}t}\left\langle j_{d}|\psi ^{b}\right\rangle
\left\langle \psi ^{b}|j_{0}\right\rangle \right\vert ^{2}.
\label{allprob_jd}
\end{eqnarray}
%\left\vert \left\langle j_{d}|\psi (t)\right\rangle \right\vert^{2}  \notag \\
The first term in the sign of absolute value is zero forever due to $\left\langle j_{d}|
\psi_{k}^{o}\right\rangle =0 $ in Eq.(\ref{evodd}). With the changes of $j_{d}-j_{0}$,
the probability deriving from the second term has larger amplitudes at the tails of its distribution,
just similar to the probability distribution of the free quantum walk induced by the interference of traveling waves. 
Unlike that, the probability deriving from the third term is mainly localized around $j_{d}-j_{0}=0$.
Compared with the third term, the second term can be neglected 
when the distance between the initial position and the defect position is not too large.
So Eq.(\ref{allprob_jd}) can be approximated as 
\begin{eqnarray}
P_{j_{d}}\approx\left\vert \left\langle \psi ^{b}|j_{d}\right\rangle \right\vert
^{2}\left\vert \left\langle \psi ^{b}|j_{0}\right\rangle \right\vert ^{2},
\label{prob_jd}
\end{eqnarray} 
which is the combined projections of the initial position state $\left\vert
j_{0}\right\rangle $ and defect position state $\left\vert
j_{d}\right\rangle $ onto the bound eigenstate $\left\vert \psi
^{b}\right\rangle $.
This approximation may be weakly depend on the defect parameter values and evolution time,
but under our choosing parameter values they are at least different from two orders of magnitude.
The height of the large sharp peak in Fig.2(a), calculating from Eq.(\ref{allprob_jd}), is 0.692427,
and the height of the smaller peak in Fig.2(b) is 0.0637546, which almost agree with the approximate results
from Eq.(\ref{prob_jd}) $P_{j_{d}}=\left(
A_{b}\frac{\gamma }{\gamma +\beta }\right) ^{4}=0.692308$ and $P_{j_{d}}=\left( A_{b}\frac{%
\gamma }{\gamma +\beta }\right) ^{2}\left( A_{b}\frac{\gamma (\alpha
+\varepsilon -\lambda _{b})}{2(\gamma +\beta )^{2}}\right) ^{2}=0.063466$, respectively.
Therefore, the spike in the probability distribution at
the defect position can be regarded as a fingerprint of this bound state, which can be
termed as eigen-localization.
When $\left\vert j_{d}-j_{0}\right\vert >1$, $P_{j_{d}}=\left( A_{b}\frac{\gamma }{\gamma
+\beta }\right) ^{2}\left( A_{b}y^{sign(1-\left\vert y\right\vert
)\left\vert j_{0}-j_{d}\right\vert }\right) ^{2}$ in  Eq.(\ref{prob_jd}) decrease
exponentially with the increase of distance $\left\vert
j_{d}-j_{0}\right\vert $ and the approximation becomes invalid.
From Fig.2(b)-(d), it is also observed that the
CTQW is largely reflected by the defect with a small probability of
transmission. Prior to encountering the defect, the CTQW is free and evolves
symmetrically in both the left and right direction. Once the part moving in
the right direction meets the defect, it will be largely reflected and move
towards the left. As a result, two envelopes appear on the left side of the
defect position and they overlap each other resulting in a complex
interference pattern, as shown in Fig.2(c) and (d).

The right panel of Fig.2 shows the CTQW probability distribution at the
defect position $j_{d}=0, 1, 2, 5$, respectively, as a function of the
defect strength $\alpha$ at $t=30$. It is shown that, although the bound
energy is less than the traveling-wave energy when $\alpha <0$ and greater
when $\alpha >0$, the probability at the defect position is symmetric about $%
\alpha =0$. That is to say, CTQW treats the single-point position defect
exactly the same regardless of it being a potential barrier or a potential
well. When the CTQW starts from the defect position, the probability
amplitude at the defect position increases monotonically with the strength
of the defect potential (see Fig.2(e)). The stronger the defect potential,
the larger the probability amplitude, with the CTQW largely localized at the
defect position. When the CTQW does not start from the defect position, the
probability at the defect position is not monotonic but rather increases
firstly and then decreases with increasing defect strength $\alpha $. It
tends to zero for the stronger defect strength.

In addition, Fig.2(a) shows that, besides a large peak
at the origin, two smaller peaks are also observed at the tails of probability 
distribution, as the same locations as the ballistic peaks of the free quantum walk.
Even when the CTQW starts from the left of the defect and it is largely reflected, as shown in
Fig.2(b)-(d), the probability distribution still has a smaller peak on the
right tail. For illustrating how a single-point position defect affect the CTQW spreads on the line,
we plot the variation of CTQW's standard deviation $\sigma =\sqrt{ \left\langle
\psi (t)\right\vert \mathbf{j}^{2}\left\vert \psi (t)\right\rangle
-\left\langle \psi (t)\right\vert \mathbf{j}\left\vert \psi
(t)\right\rangle^{2}}$ with time $t$ in Fig.3, which demonstrates predominantly a linear
relationship regardless of being localized or reflected by the defect.
\begin{figure}[th]
\begin{center}
\includegraphics [width=10cm] {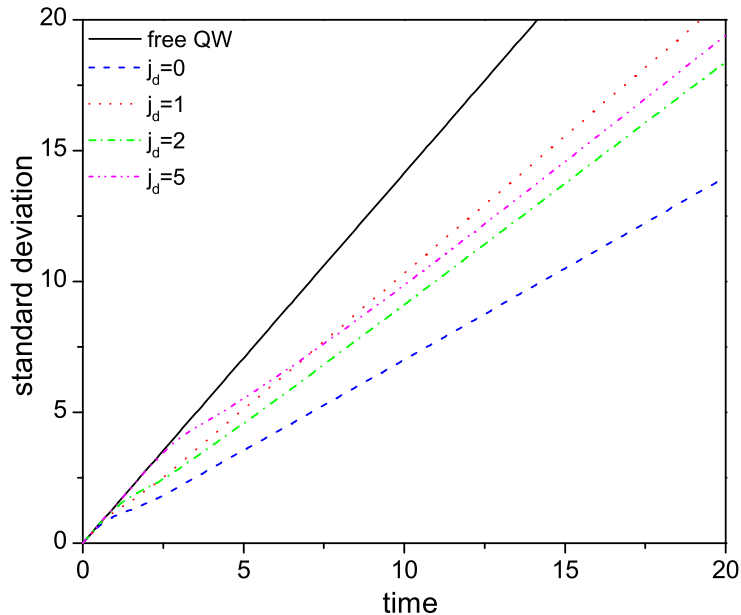}
\end{center}
\caption{The standard deviation of CTQW with a single-point position defect
as a function of time.}
\end{figure}
However, the spreading speed (given by the slope of standard deviation with
time) is dependent on the position of the defect. The appearance of
defect makes the standard deviation less than that of a defect free CTQW. As
expected, for the case $j_{d}=j_{0}$ the spreading speed is the least due to
strong localization. The pink dash-dot-dot line of $\left\vert
j_{d}-j_{0}\right\vert =5$ clearly shows that the CTQW spreads like a free
QW at the beginning, but when it encounters the defect the spreading speed
starts to decrease. In general, the larger the distance $\left\vert
j_{d}-j_{0}\right\vert $, the greater the spreading speed. As an exception,
we observe a much higher spreading speed for the case $\left\vert
j_{d}-j_{0}\right\vert =1$ (the red dotted line in Fig.3) due to the large
reflected peak at the far left end, indicating strong interference and
resonance for this special case.

\subsection{The effect of transition defect}

In this section, we focus on the effect of a single-point transition defect
on the spreading properties of CTQW. We choose the parameters $\varepsilon =2, \gamma =1$
and $\alpha=0$, the bound energy as a function of transition
defect strength is shown in Fig.4. When $\left\vert \gamma +\beta
\right\vert \leq 1$ (i.e. $-2 \leq \beta \leq 0) $, no bound eigenstate
exists, or else there are two bound states. %
\begin{figure}[th]
\begin{center}
\includegraphics [width=10cm] {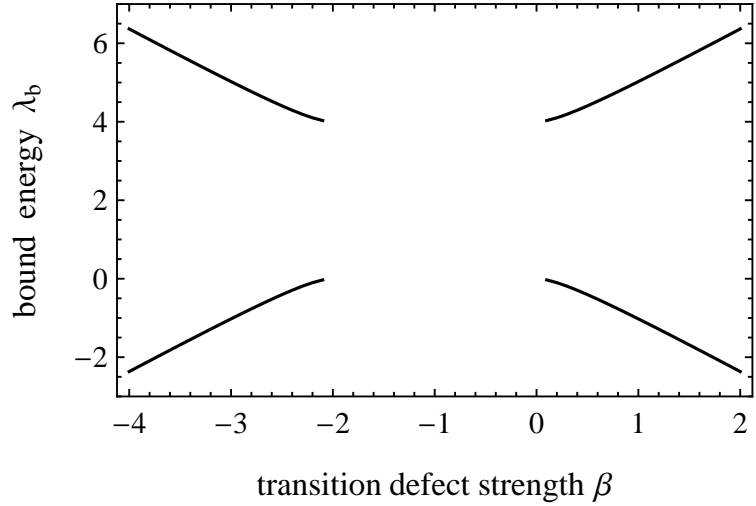}
\end{center}
\caption{The variation of bound energy with the strength of the transition
defect.}
\end{figure}
When the defect is located at the initial position ($j_{d}=j_{0}=0$), the
resulting probability distribution over the discrete position space at time $%
t=30$ is shown in Fig.5. %
\begin{figure}[th]
\begin{center}
\includegraphics [width=16cm] {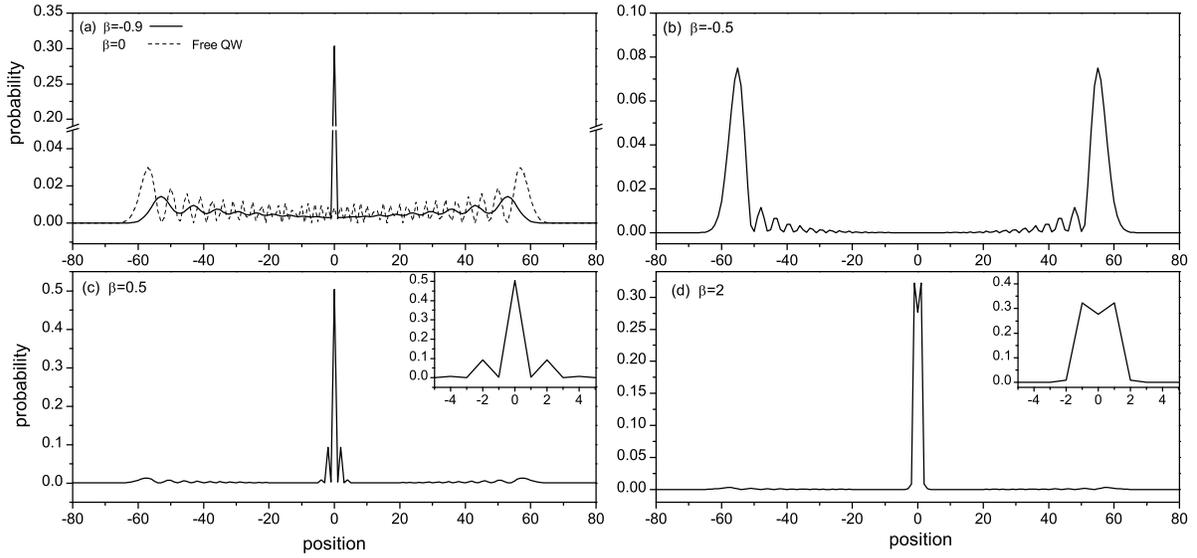}
\end{center}
\caption{The probability distribution of CTQW with a single-point transition
defect when $t=30$, $j_{d}=j_{0}=0$, and $\beta=-0.9,-0.5,0.5,2$.}
\end{figure}
Some important features to note: (1) if $(\gamma +\beta )=0$, the initial
position is disconnected from its neighbors and consequently the CTQW stays
at the initial position; (2) as $\left\vert \gamma +\beta \right\vert $
deviates slightly from zero, the residual effect of the disconnection still
shows and the probability distribution has a peak at the initial position
(see Fig.5(a)); this peak decreases with time, which distinguishes it from
the localized peak induced by eigen bound state; (3) as $\left\vert \gamma
+\beta \right\vert $ increases until it approaches 1, the CTQW spreads in a
similar way as a free QW since there is no bound state yet (see
Fig.5(b)); and (4) when $\left\vert \gamma +\beta \right\vert >1$ (e.g. $%
\beta $ = 0.5 and 2, as shown in Fig.5 (c) and (d) respectively), the
transition defect induces two bound states surrounding the defect, resulting
in a large probability in the vicinity of the defect position due to eigen-localization.

Unlike the position defect induced localization where the maximum of
probability is always at the defect position, the maximum probability
induced by a transition defect may also be at the defect neighbors (see the
insert in Fig.5(c)(d)), which is resulted by interference between the two
bound states. Neglecting the contribution from traveling eigen state,
the localization probability around defect position $\left(
j_{d}=j_{0}\right) $ can be approximately expressed by 
\begin{eqnarray}
P_{j_{d}} &=&\left( \frac{\gamma }{\gamma +\beta }\right) ^{4}\left[
A_{+}^{4}+A_{-}^{4}+2\cos [(\lambda _{+}-\lambda _{-})t]\left(
A_{+}A_{-}\right) ^{2}\right],\label{tpd1} \\
P_{j_{d}\pm 1} &=&\left( \frac{\gamma }{\gamma +\beta }\right) ^{4}\left[
A_{+}^{4}\left( \frac{\alpha +\varepsilon -\lambda _{+}}{2(\gamma +\beta )}%
\right) ^{2}+A_{-}^{4}\left( \frac{\alpha +\varepsilon -\lambda _{-}}{%
2(\gamma +\beta )}\right) ^{2}\right. \notag \\
&&\left. +2\cos [(\lambda _{+}-\lambda _{-})t]A_{+}^{2}A_{-}^{2}\frac{%
(\alpha +\varepsilon -\lambda _{+})}{2(\gamma +\beta )}\frac{(\alpha
+\varepsilon -\lambda _{-})}{2(\gamma +\beta )}\right]\label{tpd2}.
\end{eqnarray}
The last terms in the square brackets of the above equations represent the
interference between the two bound states.
The values of Eqs.(\ref{tpd1}) and (\ref{tpd2}) are approximately equal to the peak values in Fig.5(c) and (d),
fully indicating that these peaks are the eigen localization.   
In Fig.6, we plot the localized probability at defect position as a function
of the transition defect strength $\beta $ when $j_{d}=j_{0}$. %
\begin{figure}[th]
\begin{center}
\includegraphics [width=10cm] {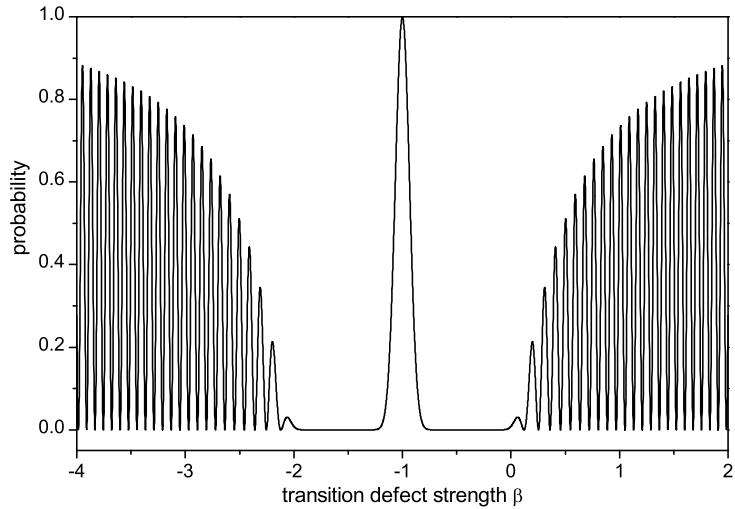}
\end{center}
\caption{The probability at defect position as a function of the transition
defect strength when $t=30$ and $j_{d}=j_{0}$.}
\end{figure}
The oscillatory behavior in the range of $\left\vert \gamma +\beta
\right\vert >1$ displays clearly the coherent effect between the two bound
states. Similar oscillation also occurs for the probabilities at the
neighbors of the defect position. When $\beta =-\gamma =-1$, complete
disconnection between the initial position and its neighbors, we have $%
P_{j_{d}}=1$. Smooth variation of $P_{j_{d}}$ with the small deviation from $%
\beta =-1$ indicates the disconnection effect remains.

The influence of a transition defect on the spread speed of CTQW is shown in
Fig.7 through the variation of its standard deviation with time. %
\begin{figure}[th]
\begin{center}
\includegraphics [width=10cm] {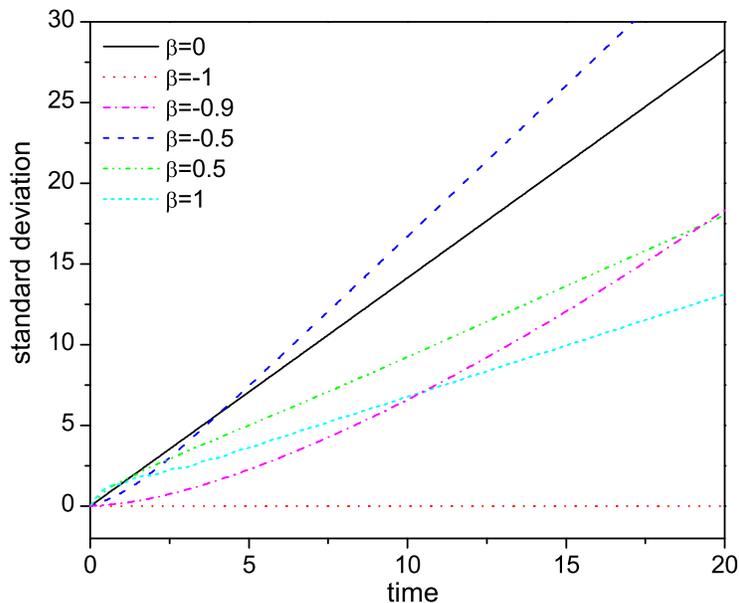}
\end{center}
\caption{The standard deviation of CTQW with a single-point transition
defect as a function of time.}
\end{figure}
One particular interesting case is $\beta =-0.5$, where the spreading speed
is significantly larger than that of a defect free CTQW, due to constructive
interference caused by the defect. In general, however, the transition
defect reduces the spreading speed due to eigen-localization and transition
defect trapping. Also, when $\left\vert \gamma +\beta \right\vert $ deviates
slightly from zero (e.g. $\beta =-0.9$), the variation of standard deviation
is clearly non-linear. This is because the residual disconnection effect
decreases with time, as the probability remaining at the initial position
decreases, and correspondingly the spreading speed increases.

When the CTQW does not start from the defect position, i.e., $j_{d}\neq j_{0}=0$,
Fig.8 presents the probability distribution at time $t=30$.
\begin{figure}[th]
\begin{center}
\includegraphics [width=16cm] {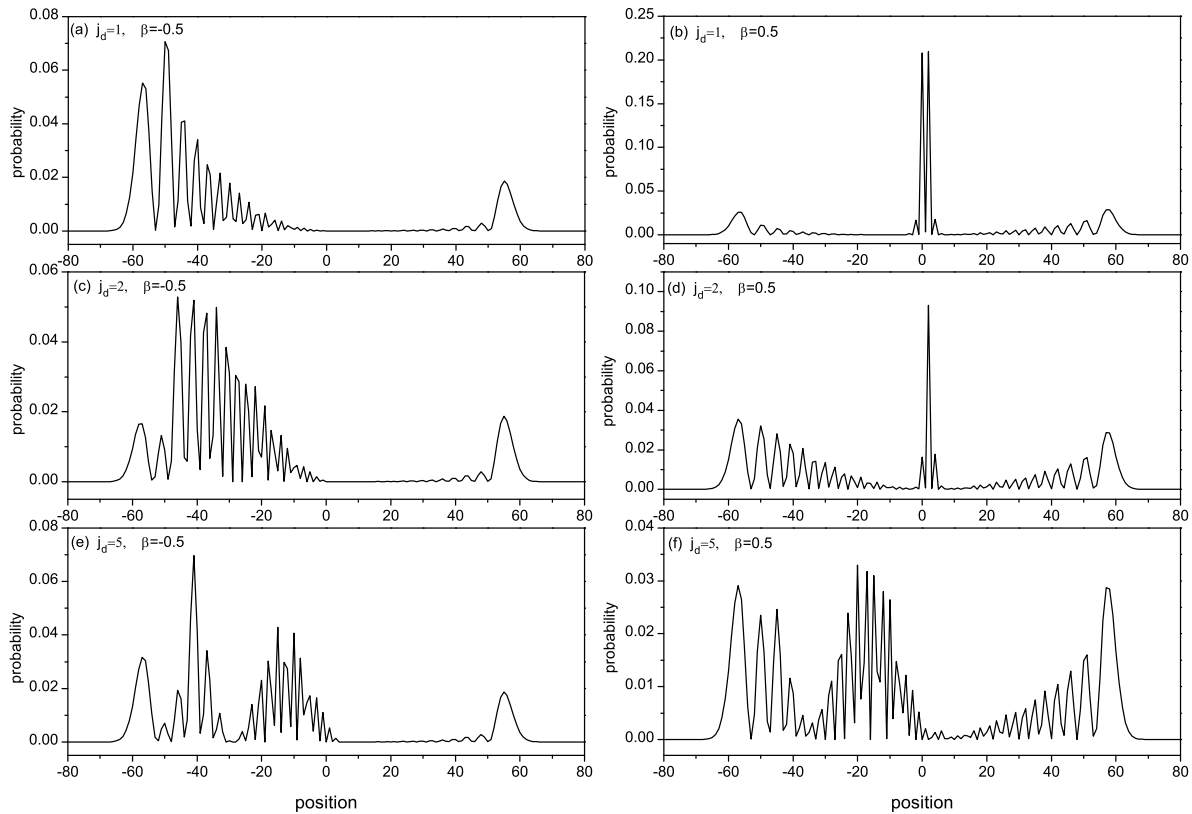}
\end{center}
\caption{The probability distribution of CTQW with a single-point transition
defect when $t=30$, $j_{0}=0$, $j_{d}=1,2,5$ and $\beta=-0.5,0.5$.}
\end{figure}
The left panel, with $\beta=-0.5$ and thus no bound state existing, shows that
the CTQW wave-packet is largely
reflected with a smaller transmission peak observed at the same locations as
the ballistic peaks of the free quantum walk. The right panel is the situation for
$\beta =0.5$, where two bound state exist. If the defect
position is the nearest to the initial position of the CTQW, $j_{d}=j_{0}+1$,
the eigen localization induced by two bound states accumulates the
probability in the vicinity of the defect position and displays strong
eigen-localisation (see Fig.8(b)). Only considering the projections of the bound eigenvectors,
we have $P_{j_{d}=1}=0.003$ and $P_{j_{d}+1=2}=P_{j_{d}-1=0}=0.209$, which is nearly equal to
the coordinate values in Fig.8(b). If the defect position goes away from the initial position,
$\left\vert j_{d}-j_{0}\right\vert >1$, the factor $y^{sign(1-\left\vert y\right\vert
)\left\vert j-j_{d}\right\vert}y^{sign(1-\left\vert y\right\vert
)\left\vert j_{d}-j_{0}\right\vert}$ in combinated projection $\left\langle j|\psi ^{b}\right\rangle
\left\langle \psi ^{b}|j_{0}\right\rangle $ makes the eigen-localization probability
decay exponentially with increasing distance $\left\vert j_{d}-j_{0}\right\vert$.

\section{Discussion}

We have introduced a new form of defects in continuous time quantum walks,
namely a single-point transition defect. A complete set of analytical
eigenvectors in position space for CTQW on the line with a single
-point position defect and a single-point transition defect is obtained.
While the system containing only a single-point position defect has one
bound state, the system possessing a single-point transition defect has
zero, one, or two bound states dependent on the transition defect
parameters. With these bound eigenstate solutions we are able to understand
the detailed dynamical properties of CTQW, including transmission,
reflection and localization. We found that the induced localization at the
defect position is determined by the combined projections of the
initial position state $\left\vert j_{0}\right\rangle $ and defect position
state $\left\vert j_{d}\right\rangle $ onto the bound eigenstate $\left\vert
\psi ^{b}\right\rangle $. Also, the coherent effect between two bound
eigenstates can be identified through the oscillating eigen localization for
the case of single-point transition defect. We present a particularly
interesting case where, due to constructive interference caused by the
defect, the spreading speed is significantly larger than that of a defect
free CTQW. This study provides another way of controlling the scattering
properties of quantum walks by introducing transition defects besides the
previously studied position defects.

This kind of eigenstate localization is different from the Anderson
localization of CTQWs. The Hamiltonian in the Anderson model are randomly chosen，
whereas the Hamiltonian under our consideration is deterministic. 
The propagation behavior for a system which exhibits Anderson
localization is that for any initial state and an arbitrary number of time
steps, and the probability to find the particle at a position is upper bounded
by an almost exponentially decaying function in the distance from its
initial position. The eigenstate localization for our model considered
depend strongly on the initial state of the quantum walker,
more precisely on the distance between the defect position and the initial position.
In fact, there are initial states such that the propagation behavior is ballistic in the
sense that the variance of the particle's position distribution grows
quadratically with time. The single-point defects in our model, as a local modification,
can be regarded as a perturbation of a translationally invariant Hamiltonian
and such perturbations generically generate bound eigenvectors.
The peak in the probability distribution, occurring around the defect,
can be understood as eigen-localization, which should be also allowed for high dimensions.

\section{Acknowledgements}

This work was supported by National Nature Science Foundation of China
(Grant No. 10974124, 11274208) and Shanxi Scholarship Council. The authors also like to acknowledge support from The University of Western Australia.

\end{document}